\documentclass{ws-ijmpa}

\newcommand{\bd}{\begin{document}}
\newcommand{\ed}{\end{document}}
\newcommand{\bc}{\begin{center}}
\newcommand{\ec}{\end{center}}
\newcommand{\beq}{\begin{eqnarray}}
\newcommand{\eeq}{\end{eqnarray}}
\newcommand{\ba}{\begin{array}}
\newcommand{\ea}{\ed{array}}
\renewcommand{\thefootnote}{\alph{footnote}}
\newcommand{\se}{\section}
\newcommand{\sse}{\subsection}

\begin{document}

\markboth{ Geng, Ho and Ng}
{CPT conserving cosmological birefringence}

%
\catchline{}{}{}{}{}
%

\title{CPT CONSERVING COSMOLOGICAL BIREFRINGENCE}

\author{ C.Q. GENG$^{1}$, S.H. HO$^1$
  and J.N. NG$^{2}$}

\address{ $^{1}$Department of Physics, National Tsing-Hua University,
Hsinchu 300, Taiwan  \\
$^2$Theory group, TRIUMF, 4004 Wesbrook Mall, Vancouver, B.C. V6T 2A3, Canada
}

\maketitle


\begin{abstract}
We demonstrate that the cosmological birefringence can arise from
$CPT$ conserving effect, originated from
  the
$CPT$-even dimension-six Chern-Simons-like term.
We show that
   a sizable
rotation polarization angle in the  data of the
cosmic microwave background radiation polarization can be induced.
\\
\\
\end{abstract}

It is well known that the polarization maps of the cosmic microwave background (CMB)
are  important tools for probing the epoch of the last
scattering directly. 
 Feng $et\ al$~\cite{r-2} have used the combined data of the
 Wilkinson Microwave Anisotropy Probe (WMAP) 
 and
 the 2003 flight of BOOMERANG (B03) 
 for the CMB polarization to  constrain
 the change of the rotation angle  $\Delta\alpha$ of the polarization,
 known as the  cosmological birefringence,
  and concluded that
a nonzero angle is mildly favored.
On the other hand, Cabella, Natoli and Silk~\cite{Silk}
 have applied a wavelet based estimator on the WMAP3 TB and EB data
 to derive a limit of $\Delta\alpha=-2.5\pm3.0$ deg, which is slightly
 tighter than that in Ref.~\cite{r-2}.
It is clear that if such rotation angle does exist, it will
 indicate an anisotropy of our Universe. It has been participated in Ref.~\cite{r-2} that the 
cosmological birefringence is a consequence of the CPT violating theory~\cite{r-3}.
 However, in this talk we will demonstrate that 
  the $CPT$ conserving interaction~\cite{Geng:2007va} could also induce the  effect.

 In Ref. \cite{r-3}, Carroll, Field and Jackiw (CFJ)
modified the Maxwell Lagrangian by adding a Chern-Simons (CS) term:
\begin{eqnarray} 
\label{lagrangian}
{\cal L}_{CS}^{CPTV}
        &  =&
        -  {1\over 2}\sqrt{g}\emph{p}_{\mu}\emph{A}_{\nu}\emph{\~{F}}^{\mu\nu}\,,
\end{eqnarray}
where
$\emph{F}_{\mu\nu}=\partial_{\mu}A_{\nu}-\partial_{\nu}A_{\mu}$ is the electromagnetic tensor,
$\emph{\~{F}}^{\mu\nu}\equiv{1\over 2}\epsilon^{\mu\nu\rho\sigma}\emph{F}_{\rho\sigma}$
is the dual electromagnetic tensor, $g$ is defined by
$g=-det(g_{\mu \nu})$, $\epsilon^{\mu\nu\rho\sigma}=g^{-1/2}e^{\mu\nu\rho\sigma}$ with
the normalization of $e^{0123}=+1$ and
 $p_{\nu}$ is a
four-vector.
 Here,  $p_{\mu}$ has been taken to be a constant  vector~\cite{r-3} or the gradient of a scalar~\cite{Carr2}.
 As shown in Table 1~\cite{CPT}, ${\cal L}_{CS}^{CPTV}$ in Eq. (\ref{lagrangian})  is $CPT$-odd \cite{coleman}. 
 In this talk, we study the possibility that the four-vector
$\emph{p}_{\mu}$ is related to a neutrino current~\cite{Geng:2007va}
\begin{eqnarray} \label{current}
\emph{j}_{\mu}&=&\bar{\nu}\gamma_{\mu}\nu\;\equiv\;(j^{0}_{\nu}, \vec{\emph{j}_{\nu}})\,,
\label{nu-current}
\end{eqnarray}
with the CS-like interaction as
\begin{eqnarray} \label{Lcs}
   {\cal L}_{CS}^{CPTC}       &=&
   -\frac{\beta}{2M^2}\sqrt{g}\emph{j}_{\mu}\emph{A}_{\nu}\emph{\~{F}}^{\mu\nu}\,,
\end{eqnarray}
 where $\beta$ is the coupling constant of
order unity and M is an undetermined new physics mass scale.
Clearly,  ${\cal L}_{CS}^{CPTC}$ in Eq. (\ref{Lcs}) 
is $P$ and $C$ odd but
$CPT$ even due to the $C$-odd vector current of $j_{\mu}$ in
Eq. (\ref{nu-current}) from Table 1.
\begin{table}[htbp]
 \tbl{
 $C$, $P$, $T$ and $CPT$ transformations.}
{\begin{tabular}{|c|c|c|c|c|} \hline
Quantity &$P$ & $C$ & $T$ &  $CPT$
\\ \hline \hline
$A^\mu$& $A_\mu$ & -$A^\mu$&$A_\mu$ & -$A^\mu$\\ \hline
$\emph{\~{F}}^{\mu\nu}$ &-$\emph{\~{F}}_{\mu\nu}$ & -$\emph{\~{F}}^{\mu\nu}$ &
 $\emph{\~{F}}_{\mu\nu}$ &$\emph{\~{F}}^{\mu\nu}$ \\ \hline
 $p^\mu$& $p_\mu$ & $p^\mu$&$p_\mu$ & $p^\mu$\\ \hline
 $J^\mu$& $J_\mu$ & -$J^\mu$&$J_\mu$ & -$J^\mu$
\\ \hline
\end{tabular}
\label{Table1}}
\end{table}
Furthermore, as
${\cal L}_{CS}^{CPTC}$  is a
dimension-6 operator, it must be suppressed by two powers of
the mass scale $M$.
Note that $\vec{\emph{j}_{\nu}}$
is the neutrino flux density and $j^{0}_{\nu}$ is the number density difference between neutrinos and anti-neutrinos, given by
\begin{eqnarray}
j^{0}_{\nu}&=&\Delta n_{\nu}\;\equiv\; n_{\nu}-n_{\bar{\nu}}\,,
\label{Delta-n}
\end{eqnarray}
where $n_{\nu(\bar{\nu})}$ represents the neutrino (anti-neutrino)
number density. 

As we pointed out in Ref. \cite{Geng:2007va}, the interaction (\ref{lagrangian}) is gauge invariant and the current in Eq. (\ref{current}) takes the form 
\beq
 \emph{j}_{\mu}=\bigg{(} (n_{\nu}-n_{\bar{\nu}}), \vec{0} \bigg{)}
\eeq
for a comoving observer.
 However, we remark that  Eq. (\ref{Lcs}) is not formly gauge invariance.
 In general, to maintain the gauge invariance,
 we have to introduce the St$\ddot{u}$ckelberg field \cite{Geng:2007va,jackiw}.
Moreover, the existence of a non-zero component $j_{\nu}^{0}$ in Eq. (\ref{Delta-n}) would violate Lorentz invariance \cite{r-3}.

Following
 Ref. \cite{r-3},
 the change  in the
position angle  of the polarization plane $\Delta\alpha$
 at redshift $z$
 due to our Chern-Simons-like term
  is given by
\begin{equation} \label{angle-1}
\Delta\alpha=\frac{1}{2}\frac{\beta}{M^2}\int \Delta n_{\nu}(t)
\frac{\textit{d}t}{R(t)}\,,
\end{equation}
where
to describe a
flat, homogeneous and isotropic universe,
we use the Robertson-Walker metric
\begin{equation}
ds^2=-dt^2 +R^2(t)\;d\textbf{x}^2\,,
\end{equation}
 with R being the scale factor.
To find out $\Delta\alpha$, we need to know the neutrino asymmetry
in our Universe, which is strongly constrained by the BBN
abundance of $^4$He. It is known that for a lepton flavor, the
asymmetry is given by: \cite{r-6,r-6r}
\begin{eqnarray} \label{asym-1}
\eta_{\ell}&=&\frac{n_{\ell}-n_{\bar{\ell}}}{n_{\gamma}}\;=\;\frac{1}{12\zeta(3)}
\left(\frac{T_{\ell}}{T_{\gamma}}\right)^3 (\pi^2
\xi_{\ell}+\xi_{\ell}^3)\,, \end{eqnarray} where $n_{i}\ (i=\ell,\gamma)$
are the $\ell$ flavor lepton and photon number densities, $T_{i}$
are the corresponding temperatures and
$\xi_{\ell}\equiv\mu_{\ell}/T_{\ell}$ is the degeneracy parameter, respectively.

As noted
by Serpico and Raffelt \cite{r-6}, the lepton
asymmetry in our Universe resides in neutrinos because of the
charge neutrality, while the neutrino number asymmetry depends
only on the electron-neutrino degeneracy parameter $\xi_{\nu_{e}}$
since neutrinos reach approximate chemical equilibrium before BBN
\cite{r-11}. From Eq. (\ref{asym-1}), the neutrino number
asymmetry for a lightest and relativistic, say, electron neutrino
is then given by
\cite{r-6,r-6r,r-6more}:
\begin{equation} \label{asym-2}
\eta_{\nu_e}\simeq 0.249 \xi_{\nu_{e}}
\end{equation}
where we have assumed $(T_{\nu_{e}}/T_{\gamma})^3=4/11$.
Note that the current
bound on the degeneracy parameter is $-0.046<\xi_{\nu_{e}}<0.072$
for a $2\sigma$ range of the baryon asymmetry~\cite{r-6,r-6r}. 
 From Eqs. (\ref{Delta-n}),  (\ref{asym-1}) and (\ref{asym-2}),
we obtain
\begin{eqnarray} \label{asym-3}
\Delta n_{\nu}&
         \simeq & 0.061\xi_{\nu_{e}}T_{\gamma}^3\,,
\end{eqnarray}
where we have used $n_{\gamma}=2\zeta(3)/\pi^2 \  T_{\gamma}^3$.

Consequently,  Eq. ({\ref{angle-1}) becomes
\begin{eqnarray}
 \label{angle-2}
\Delta\alpha &=&
\frac{\beta}{M^2} 0.030 \xi_{\nu_{e}} (T_{\gamma}^{\prime})^3 \int_{0}^{z_*} (1+z)^3
 \frac{\textit{d}z}{H(z)}\,,
\end{eqnarray}
where $T_{\gamma}^{\prime}$ is the photon temperature in the present time,
$H(z)=H_0(1+z)^{3/2}$ in a flat and matter-dominated Universe and
 $H_0=2.1332 \times 10 ^{-42}h\ GeV$ is the Hubble constant
with  $h\simeq 0.7$ at the present.
We note that
as the rotation angle in Eq. (\ref{angle-2}) is mainly generated at the last scattering surface, there is no rotation of the large-scale CMB polarization which is generated by reionization at $z\sim 10$. However,
due to the accuracy level of current CMB polarization data, we
  have assumed a constant rotation angle over all angular scales.
 Finally, by taking $1+z_*=(1+z)_{decoupling}\simeq 1100$ at the photon
decoupling and $T_{\gamma}^{\prime}=2.73K$,
we obtain
 \begin{eqnarray} \label{angle-4} \Delta\alpha &\simeq& 4.2\times
10^{-2}\beta\left({\xi_{\nu_{e}}\over 0.001}\right)
\left({10\,TeV\over M}\right)^{2}\,. 
\end{eqnarray} 
As an illustration,  by taking $\beta\sim 1$, $M\sim 10\ TeV$ and
$\xi_{\nu_{e}}\sim \pm 10^{-3}$, we get $\Delta \alpha
\sim\pm4\times 10^{-2}$, which could explain the results in Ref.
\cite{r-2}. We note that a sizable $\Delta \alpha$ could be
still conceivable even if the neutrino asymmetry is small. In that case,
the scale parameter $M$ has to be smaller.


In summary, 
we have proposed a new type of effective interactions  \cite{Geng:2007va}
in terms of the
$CPT$-even dimension-six CS-like term
 to generate the cosmological birefringence without violating the CPT symmetry.
To induce a sizable
rotation polarization angle in the CMB data,  a non-zero
neutrino number asymmetry is needed.
We remark that
the Planck Surveyor \cite{Planck} will reach a sensitivity  \cite{Lue,WTNi-r} of $\Delta \alpha$
at levels of $10^{-2}-10^{-3}$, while a dedicated future experiment on
the cosmic microwave background radiation polarization would reach
$10^{-5}-10^{-6}$ $\Delta \alpha$-sensitivity~\cite{WTNi-r}.

Finally, we would like to mention that the CPT-even CS-like interaction in
Eq. (\ref{Lcs})
could also be used to understand
 the origin of baryogenesis \cite{Bamba:2007hf} when the neutrino current and electromagnetic field are  replaced by
the baryon current and  hypermagnetirc field. 
The new interaction could effectively produce an energy split 
between baryons and anti-baryons if the hypermagnetic helicity exists. As the sphaleron effect in the SM is working effectively, which provides a source of baryon number violation processes, this energy split results in 
the observed baryon asymmetry in our Universe. 

\section*{Acknowledgments}
We would like to thank  Professor Chun Liu for organizing this
wonderful conference.
Two of us (CQG and SHH) would like thank the financial support from ITP and KITPC during the conference.
This work is supported in part by
the Natural Science and Engineering Council of Canada
and
the National Science Council of
R.O.C. under Contract \#: NSC-95-2112-M-007-059-MY3.


\end{document}